\documentclass[12pt]{article}

\def\rdots{\mathinner{\mkern1mu\raise1pt\vbox{\kern1pt\hbox{.}}\mkern2mu
   \raise4pt\hbox{.}\mkern2mu\raise7pt\hbox{.}\mkern1mu}}
\newcommand{\Z}{{\rm Z\kern-.35em Z}}
\newcommand{\bP}{{\rm I\kern-.15em P}}
\newcommand{\Q}{\kern.3em\rule{.07em}{.65em}\kern-.3em{\rm Q}}
\newcommand{\R}{{\rm I\kern-.15em R}}
\newcommand{\h}{{\rm I\kern-.15em H}}
\newcommand{\C}{\kern.3em\rule{.07em}{.65em}\kern-.3em{\rm C}}
\newcommand{\T}{{\rm T\kern-.35em T}}

\newcommand{\be}{\begin{equation}}
\newcommand{\ee}{\end{equation}}

\newcommand{\ve}{\varepsilon}

\newcommand{\ga}{\gamma}

\newcommand{\pa}{\partial}

\newcommand{\Om}{\Omega}
\newcommand{\de}{\delta}

\newcommand{\al}{\alpha}
\newcommand{\nn}{\nonumber}

\begin{document}
  
\openup 1.5\jot
\centerline{On Schwarzschild-Like Solutions in Curvature-Quadratic Gravity}

\vspace{1in}
\centerline{Paul Federbush}
\centerline{Department of Mathematics}
\centerline{University of Michigan}
\centerline{Ann Arbor, MI 48109-1109}
\centerline{(pfed@math.lsa.umich.edu)}

\vspace{1in}

\centerline{\underline{Abstract}}

Partial results are obtained for Schwarzschild-like solutions in a gravity theory with action density $c(-g)^{1/2} [R^2_{ik} + bR^2]$.  A seven parameter family of implicit solutions is found.  A number of explicit solutions are also exhibited.

\vfill\eject

In a previous paper, [1], we have proposed the curvature-quadratic action
\be	c \int d^4x (-g)^{1/2} \left[ R^{ik} R_{ik} + bR^2 \right]	\ee
as the basis for quantum gravity.  If $b = - \frac 1 3$ then this action is of the conformal Weyl theory.  Classical solutions of the Euler-Lagrange equations for this conformal theory have been studied by Mannheim and Kazanas, in particular for Schwarzsc
hild-like solutions, [2].  In this paper we study the problem of such solutions for the action (1) and arbitrary value of $b$.  (In [3] there are some results on the linearized equations.)

We thus seek static spherically symmetric metrics.  By a coordinate change these may be put in form
\be
ds^2 = d^2(r) \left[ -c(r)dt^2 + \frac 1{c(r)} (dr)^2 + r^2 \; d\Om \right]	
\ee
(as shown in [2]).  For the Einstein action
\[	c \int d^4x(-g)^{1/2} \ R	\]
there is the solution of the form (2) as follows:
\be	d(r) = 1, \ \ c(r) = 1 + \frac \al r \ .	\ee
In the conformal Weyl theory, the solution metrics are of the form:
\be	c(r) = 1 - \beta(2-3\beta\gamma) \frac 1 r - 3\beta\ga + \ga r - kr^2    \ee
(See [2].)  In this case, of course, $d(r)$ is arbitrary.

For the action (1), general $b$, our results are partial.  There are several explicit solutions:

\noindent
\underline{explicit solution 1}:
\be	c(r) = 1+ \frac \gamma r, \ \ d(r) = 1+ (\frac \al \gamma) \ \ell n (1 +  \frac \gamma r) + \beta	\ee

\noindent
\underline{explicit solution 2}:
$$	c(r) = 1, \ \ d(r) = 1 + \frac \al r + \beta  
  \eqno(5')    $$
This is a limit of the solution in equation (5).

\noindent
\underline{explicit solution 3}:
\be	c(r) = 1 + \frac \al r + \beta r^2 , \ d(r) = 1 + \ga \ .    \ee
This is the well-known de Sitter solution.

\noindent
\underline{explicit solution 4}:
\be 
c(r) = 1 + \al^2 r^2, \ d(r) = \frac \beta {(\al r)} \arctan (\al r)
\ee

\noindent
\underline{explicit solution 5}:
\begin{eqnarray}
c(r) &=& 1 - 2 \alpha r + 4 \alpha^2 r^2 \nonumber \\
d(r) &=& \beta\left( \frac 1 {2\alpha r} \right) \frac 2 3 \sqrt{3} \arctan \left( \frac{\sqrt{3}} {3} (4 \alpha r - 1) \right) 
\end{eqnarray}
 Beyond these, there is the formal solution.

\be	(c(r), d(r)) = (1,1) + \sum^\infty_{n=1} t^n \vec{u}_n	\ee
$$	\vec{u}_1 = \sum^7_{j=1}  a_j \vec{\phi}_j	\eqno(9')  $$
$$	\vec{\phi}_1 = (0, 1) \eqno(10a)   $$
$$	\vec{\phi}_2 = (r^2, 0) \eqno(10b)   $$
$$	\vec{\phi}_3 = (0, r^2) \eqno(10c)   $$
$$	\vec{\phi}_4 = (\frac 1 r, 0) \eqno(10d)   $$
$$	\vec{\phi}_5 = (0, \frac 1 r) \eqno(10e)   $$
$$	\vec{\phi}_6 = (-2r, r) \eqno(10f)   $$
$$	\vec{\phi}_7 = \left( -4r ln \; r, 2r \; ln \; r - \frac 1 2 \frac{3+8b}{3b+1} \ r \right) \eqno(10g)   $$
The $\vec{u}_i$ may be found inductively, each is a finite sum of terms, polynomial in the $a_i, r, 1/r, ln\; r$.  The sum in (9) is essentially a perturbation expansion.  The convergence may be described as follows:  for each finite interval, $\al < r < 
\beta$, and choice of the $a_i$, there is an $\ve$ such that the series in (9) converges if $0\le |t| < \ve$, and represents a solution of the Euler-Lagrange equations.

The Euler-Lagrange equations were derived as follows.  A metric of the form

\setcounter{equation}{10}

\be	ds^2 = \left(-d^2(r) c(r) + B(r) \right)dt^2 + \left(-d^2(r) + A(r) \right)r^2 d\Om
+ \left( \frac{d^2(r)}{c(r)} + A(r) + r^2 \; C(r) \right) dr^2   \ee
was substituted into the action (1); and three Euler-Lagrange equations
\begin{eqnarray}
X &=& 0 \nn \\
Y &=& 0  \\
Z &=& 0 \nn 
\end{eqnarray}
were obtained by requiring the action to be invariant under variations, with respect to $A(r), B(r)$ and $C(r)$ respectively (evaluated at $A=B=C=0$).  This was done using Maple software.  Each of the equations  in (12) has approximately 500 terms, the al
gebra could not be done by hand.  There are thus three equations to be satisfied by the two functions $c(r)$ and $d(r)$.  It was then checked (in Maple) that (5), (6), (7), and (8) satisfied the equations (12).  (The exact form in which the variations $A,
 B$, and $C$ were put in (11) was a largely arbitrary choice.)

Under a coordinate change
\be	r  \longrightarrow  r + \ve \phi(r)	\ee
the action (1) remains invariant, but the $A, B$, and $C$ change as follows:
\begin{eqnarray}
\de A &=& \ve \left[ \phi \; \frac{\pa (d^2(r))}{\pa r} + 2d^2(r) \frac 1 r \; \phi \right] \\
\de B &=& \ve \; \phi \frac \pa{\pa r} \Big( -d^2(r) c(r) \Big) \\
\de C &=& \ve \bigg[ \phi \; \frac 1{r^2} \frac \pa{\pa r}  \left(\frac{d^2(r)}{c(r)} \right) + 2 \frac{1}{r^2} \frac{\pa \phi}{\pa r} \ \frac{d^2(r)}{c(r)} \nn \\
&-& \frac 1{r^2} \left( \phi \frac {\pa}{\pa r} ( d^2(r)) + 2 d^2(r) \frac 1 r \ \phi \right) \bigg] \ .
\end{eqnarray}
This leads to the relation between $X, \ Y, \  Z$ of (12) as follows:
\begin{eqnarray}
\bigg( \frac d{dr} ( d^2(r) ) &+& 2 d^2(r) \frac 1 r \bigg) X + \frac d {dr} 
\big( -d^2(r)c(r) \big) Y \nn \\
+ \bigg( \frac 1{r^2} \frac d{dr} \left( \frac{d^2(r)}{c(r)} \right) &-&  \frac 1 {r^2} \frac d{dr}  \big(d^2(r)\big) - \frac 2{r^3} \ d^2(r) \bigg) Z \\
&-& \frac d {dr} \left(2 \ \frac{d^2(r)}{c(r)} \ \frac 1{r^2} Z \right) = 0 . \nn
\end{eqnarray}

We can see from (17) that if the last two equations of (12) hold, the first automatically is satisfied.  The terms in (10) arise as a complete set of seven linearly independent solutions of the linear terms in the equations $Y = 0, \ Z = 0$, the first of 
these is a fourth order differential equation, the second a third order.  The linear portions of $Y=0$ and $Z = 0$ are easily verified to be a non-singular linear system (that of $X=0$ and $Z = 0$ would not be).  This ensures that during the iterative con
struction of the $\vec{u}_i$ no compatibility requirements are met.  i.e. once $\vec{u}_1, ..., \vec{u}_N$ have been found, the construction of $\vec{u}_{N+1}$ does not impose restrictions on the $\vec{u}_1, ..., \vec{u}_N$ already developed.

It is of course of interest to see if other explicit solutions may be found. 

\underline{Acknowledgment}:  I would like to thank Professor Hans-Juergen Schmidt for helpful comments.

\vfill\eject

\centerline{\underline{References}}

\begin{itemize}
\item[[1]] P. Federbush, ``A Speculative Approach to Quantum Gravity", Symposium in Honor of Eyvind H. Wichmann, University of California, Berkeley, June 1999.
\item[[2]] P.D. Mannheim and D. Kazanas, ``Exact Vacuum Solution to Conformal Weyl Gravity and Galactic Rotation Curves", {\it Astrophys. J.} {\bf 342} 635 (89).
\item[[3]]  M. Roberts, ``Galactic Rotation Curves and Quadratic Lagrangians", {\it Monthly Notices of the Royal Astronomical Society}, Vol. 249, p. 339 (91).
\end{itemize}

\end{document}